\documentclass{article}
\usepackage[T1]{fontenc}
\usepackage{lmodern}
\usepackage[preprint]{spconf}
\usepackage{amsmath,graphicx}

\title{3-KEY-INPUT: EXPLORING THE THEORETICAL MINIMUM KEYS FOR TEXT ENTRY}
\name{Naoki Kimura}
\address{LY Corporation, Tokyo, Japan\\ naokkimu@lycorp.co.jp}
\toappear{Published in: ICASSP 2026 - 2026 IEEE International Conference on Acoustics, Speech and Signal Processing (ICASSP)}

\begin{document}
\ninept
\maketitle
\begin{abstract}
How far can we reduce the number of physical keys if we endow an ambiguous keyboard with modern language models? Fewer keys increase hardware design freedom in constrained settings such as assistive devices and mobile form factors. This paper systematically evaluates text entry systems using 2--5 physical keys combined with language-model-based disambiguation. On a 300-sentence English corpus (100 sentences each for Business / Conversational / Technical), we compare key counts (2--5), letter-to-key mappings (layout-based / frequency-based / intentionally worst-case), and decoders (Trie-only, GPT-2 beam search, GPT-4o selection). We find that 3 keys + GPT-4o achieves character error rate (CER) 9.46\% and word error rate (WER) 12.20\%, reducing CER by 59\% relative to 2 keys (CER 23.3\%). At 3 keys, the key-stream entropy is 1.54 bits/char; while increasing to 5 keys improves accuracy (CER 5.4\%), the marginal gains diminish. Mapping choice has a small impact under standard designs ($\Delta$CER $< 0.5$\,pp), and even an intentionally worst mapping degrades CER by only +0.5\,pp, whereas Technical sentences yield roughly twice the error rate of Business. These results suggest that, in our evaluated offline setting under a strong LM prior, 3 keys are a practical minimum for general English.
\end{abstract}

\section{Introduction}

How far can we minimize the number of keys if we give an ambiguous keyboard the power of modern language models? An \emph{ambiguous keyboard} reduces the number of physical keys by assigning multiple letters to a single key; T9 input on feature phones (9 keys) is a well-known example. The fewer keys we need, the greater the design freedom for devices in constrained settings such as assistive technology and mobile form factors.

Prior work has mainly approached this topic as text entry research in Human--Computer Interaction. Text entry research typically assumes that a person types the text by hand and edits it by hand, and that the produced text is delivered as-is to another person. Therefore, it typically does not tolerate residual errors in the final text or delayed feedback. As a result, using computationally heavy language models to probe the lower bound of key count has not been a common motivation.

However, the emergence of large language models (LLMs) is changing this assumption. Input to LLM chat does not necessarily require perfect correctness, and even when sending messages to people or publishing text, LLM-centered proofreading is becoming common. By relaxing the fixed premises of traditional text entry, and by assuming an LLM in the loop and stronger compute on mobile devices, there is room to redesign text entry systems.

From an information-theoretic perspective, English contains roughly 1.1 bits/character, while a 2-key interface provides 1.0 bit/press and a 3-key interface provides $\log_2 3 \approx 1.58$ bits/press.

This paper temporarily removes the constraints of immediacy and device compute limits, and asks: how close can we get to the theoretical limit by leveraging today's language modeling capabilities? The scope of this paper is \emph{after removing immediacy and device compute constraints, how much ambiguity can be compensated by language-model-based disambiguation?} Conversely, although this paper is related to text entry research, it is out of scope to cover current hardware constraints, recovery from input errors in realistic UX, the actual cognitive load, input speed, learning curves, and user studies tied to those topics.

In controlled evaluations on a 300-sentence corpus spanning business, conversational, and technical text, we find that 3 keys + GPT-4o achieves CER 9.46\% (WER 12.20\%), reducing CER by 59\% relative to 2 keys (CER 23.3\%). At 3 keys, the key-stream entropy is 1.54 bits/char; while increasing to 5 keys improves accuracy (CER 5.4\%), the marginal gains diminish. Mapping choice has a small impact under standard designs ($\Delta$CER $< 0.5$\,pp), and even an intentionally worst mapping degrades CER by only +0.5\,pp, whereas Technical sentences yield roughly twice the error rate of Business.

Prior work reduces keys or leverages prediction but typically avoids the level of ambiguity induced by 2--3 keys. Classical 9--12 key methods such as T9 emphasize lexicon and layout \cite{oney2013zoomboard,chen2014swipeboard}, while recent small-screen techniques (ZoomBoard, SwipeBoard), hardware variants (ForceBoard, DualKey), and one-switch timing interfaces (Nomon) focus on spatial or temporal precision rather than pushing ambiguity to the extreme \cite{oney2013zoomboard,chen2014swipeboard,zhong2018forceboard,gupta2016dualkey,bonaker2021nomon}. Input decoding has evolved from statistical touch/noise models (e.g., VelociTap) to neural decoders in production keyboards and general language models (LMs) \cite{vertanen2015velocitap,zhang2019neural,devlin2019bert,brown2020language}. Concurrently, compression results highlight the equivalence between language modeling and coding efficiency \cite{ozturk2024mlmcompress,schmidhuber1996neural,nesse2023gptzip}. Our contribution is to combine extreme key reduction (2--5 keys) with modern LM-based disambiguation, mapping $\ge 26$ letters onto few keys and resolving ambiguity by sentence context.

We implement a modular prototype: letter-to-key mappings (layout-based, frequency-balanced, and an adversarial ``worst'' map), trie-based candidate generation with diversity filtering, and sentence-level LM selection. Across keys, the 2$\to$3 transition yields the largest accuracy gain; beyond three, returns shrink. Reasonable mapping strategies perform near-identically at three keys ($\Delta$CER $< 0.5$ percentage points), implying designers may optimize for ergonomics or hardware constraints rather than micro-layout. The adversarial clustering of frequent letters degrades CER modestly ($+\approx 0.5$ percentage points), suggesting only simple cautions suffice. Domain shift remains the principal challenge: technical text exhibits roughly 2$\times$ the CER of business text due to specialized lexicons and acronym handling, indicating a need for domain lexicons or constrained decoding.

These findings have design implications. For augmentative and alternative communication (AAC), three keys strike a balance between ambiguity and recoverability for users operating eye, head, electromyography (EMG), or sip-and-puff switches. Two keys are viable for fixed-phrase menus or tightly constrained domains but are too error-prone for general free-form text. Where technical vocabulary is common, designers can either inject domain lexicons or expose a 4-key mode to reduce candidate branching while preserving a compact device. Because mapping differences are small under strong language priors, implementations should prioritize reliable actuation, low false activations, confirmation/undo flows, and on-device privacy for personal lexicons. In wearables such as smartwatches and smart glasses, three keys map naturally onto tap/press/scroll or three touch zones, enabling eyes-up, one-handed input with LM-backed disambiguation and phone-assisted rescoring.

In summary, we provide (1) an empirical lower bound suggesting that three keys can be a practical minimum for general English in our evaluated offline setting under contemporary LMs, (2) evidence of mapping robustness under strong language priors, (3) quantification of LM gains over statistical baselines under extreme ambiguity, and (4) a design lens that connects entropy, key count, and LM capacity to concrete AAC and wearable interfaces. By situating our results within information-theoretic limits and modern decoding, we motivate practical 3-key systems in this offline setting and point to domain-tuned, low-latency decoders for future real-time deployment. To reduce risks of pretraining overlap, we use a proprietary 300-sentence corpus and verify non-overlap via hashing and embedding screening (Sec.~\ref{sec:leak}).

\section{Theory: Capacity vs. Conditional Entropy}

Let $X$ be the character sequence and $C$ the linguistic context. We compare the requirement $H(X\mid C)$ to the device capacity of an $N$-key interface with keystrokes per character (KSPC), where KSPC$=1$ in our setup; this supplies at most $\log_2 N$ bits/character. If $\log_2 N < H(X\mid C)$, zero error is information-theoretically impossible.

Let $K$ denote the key sequence and define the observation $Z=(K,C)$. A Fano-type bound \cite{cover2006elements} yields
\[
H(X\mid Z)\le h(P_e)+P_e\,\log(|\mathcal{X}|-1),
\]
so decreasing $H(X\mid C)$ (stronger LM/context) or increasing $\log_2 N$ reduces the achievable error rate $P_e$.

Empirical efficiency is reported as bits/character
\[
\mathrm{bits/character} = \frac{H(K)}{\mathrm{char/press}}, \quad H(K) = -\sum_{k=1}^{N} p(k)\log_2 p(k),
\]
with $\mathrm{char/press}{=}1$ in our setup. Non-uniform key usage makes $\mathrm{bits/character} \le \log_2 N$ (e.g., $\approx 1.54$ for 3-key vs. $\log_2 3 \approx 1.585$).

\section{System Overview}
\noindent\emph{Goal.} A modular prototype that maps ambiguous $N$-key sequences to plausible text via candidate generation and LM-based selection. The system is for theoretical evaluation, not real-time deployment.

\subsection{Architecture}
\noindent\emph{Key mapping.} Assign the 26 letters to $N \in \{2,3,4,5\}$ keys. Methods: (a) layout-based groupings of QWERTY (left/middle/right or rows), (b) frequency-balanced partitions to balance key entropy, and (c) intentionally ``worst-case'' mappings that concentrate frequent letters on the same key. (b) and (c) are designed as advantageous vs.\ disadvantageous mappings for theoretical evaluation. (a) assumes that users who learned QWERTY can leverage existing knowledge and muscle memory about left/right (or row-wise) letter groupings.\par
\noindent\emph{Candidate generation.} A trie over a frequency-annotated lexicon enumerates word candidates compatible with each key sequence. Diversity-aware pruning prevents near-duplicate candidates.\par
\noindent\emph{Sentence prediction.} Three stages: (a) frequency baseline, (b) GPT-2 beam search, (c) GPT-4o selection using bidirectional context and a ranked candidate set.\par
\noindent\emph{Design choices.} Sentence-level processing, offline metrics, and LM API usage, focusing on isolating the effects of key count, mapping, and LM strength.

\begin{figure*}[t]
\centering
\includegraphics[width=\textwidth]{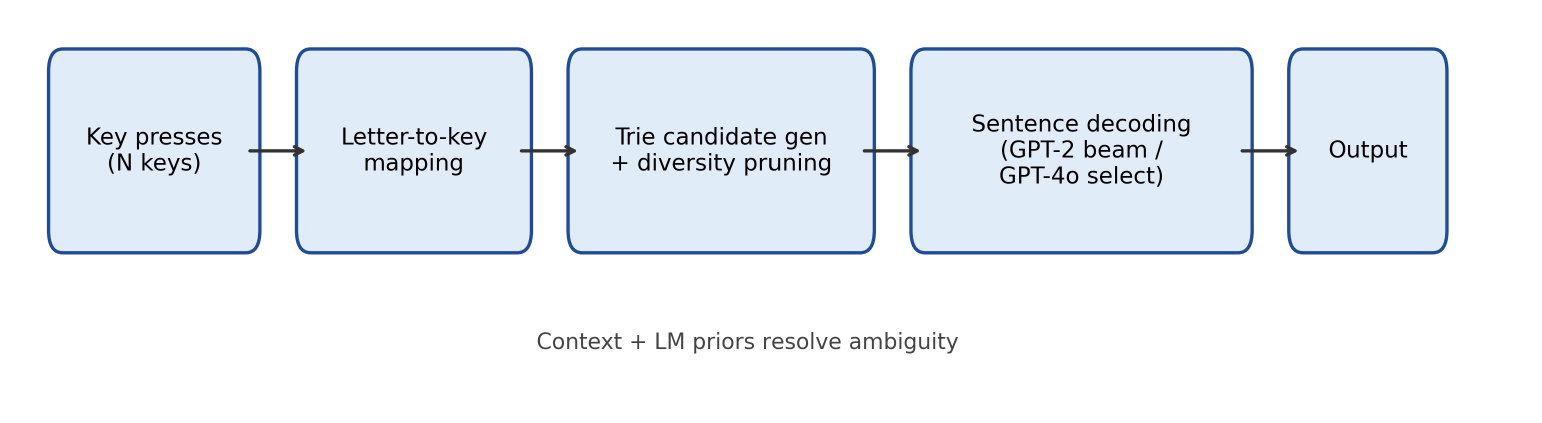}
\caption{System overview: N-key inputs are mapped to letter groups, a trie enumerates candidates with diversity pruning, and sentence-level decoding (GPT-2 beam / GPT-4o selection) chooses the output.}
\label{fig:system}
\end{figure*}

\subsection{Key Mapping}
\noindent\emph{Strategies.} \emph{Layout-based:} Partition QWERTY into left/middle/right groups or by rows. \emph{Frequency-based:} High/medium/low frequency letters distributed to balance key entropy; example for 3-key: \{e,t,a,o,i,n,s,h\}, \{r,d,l,c,u,m,w,f,g\}, \{y,p,b,v,k,j,x,q,z\}. \emph{Worst-case:} Concentrate frequent letters on the same key.

\noindent\emph{Information capacity.} Theoretical per-press capacity is $\log_2 N$ bits; thus 2-key: 1.00, 3-key: 1.58, 5-key: 2.32. Frequency-based 3-key mappings achieve $\approx 1.56$ bits/character empirically due to English letter frequencies. The specific 3-key assignments we evaluate are:
\begin{table*}[t]
\centering
\small
\setlength{\tabcolsep}{12pt}\renewcommand{\arraystretch}{1.35}
\begin{tabular}{@{}l p{4.4cm} p{4.4cm} p{4.4cm}@{}}
\hline
\textbf{Mapping} & \textbf{Key 1} & \textbf{Key 2} & \textbf{Key 3} \\
\hline
Left/Middle/Right &
a, q, z, w, s, x &
e, d, c, r, f, v, t, g, b &
y, h, n, u, j, m, i, k, o, l, p \\
Top/Middle/Bottom &
q, w, e, r, t, y, u, i, o, p &
a, s, d, f, g, h, j, k, l &
z, x, c, v, b, n, m \\
Frequency-based &
e, t, a, o, i, n, s, h &
r, d, l, c, u, m, w, f, g &
y, p, b, v, k, j, x, q, z \\
Adversarial worst &
e, t, a, o, i, n, s, h, r &
d, l, u, m, w, f, c, g, y &
p, b, v, k, j, x, q, z \\
\hline
\end{tabular}
\caption{Letter-to-key groupings for 3-key mappings. 2/4/5 keys follow spatial/frequency splits.}
\label{tab:maps-3key}
\end{table*}

\subsection{Candidate Generation}
A trie gives $O(L)$ lookup for word length $L$ and shares prefixes across candidates. Each node stores children, an end-of-word flag, and frequency. Ambiguous lookup takes a per-position character set induced by the key mapping and traverses feasible paths. Diversity filtering keeps varied morphological patterns rather than returning only high-frequency near-duplicates.

\subsection{Sentence Prediction}
\noindent\emph{Baselines.} \emph{Freq-only:} Greedy per-position most frequent candidate. \emph{LM beam:} GPT-2 scoring with caching across beams to reduce recomputation. \emph{GPT-4o selection:} Provide previous/following context and top candidates per slot; use low temperature for stability.

\noindent\textit{LM settings.} For GPT-4o selection we use temperature $=0.2$ for stability, present candidates in frequency order with prior/following context when available, batch requests to respect rate limits, and employ robust fallbacks if API calls fail or parsing errors occur.

\subsection{Optimizations and Edge Cases}
\noindent\emph{Edge cases.} Single-character words handled with priors for \textit{a}/\textit{I}. Robust fallbacks used if parsing fails. Batched API calls mitigate rate limits. \emph{Performance.} Prefix compression and frequency thresholds reduce trie memory. Caching accelerates LM beam search.

\section{Results and Discussion}

\subsection{Setup}
We report character error rate (CER), word error rate (WER), BLEU score~\cite{papineni2002bleu} (where informative), BERTScore F1~\cite{zhang2019bertscore} (model: \texttt{roberta-large}; baseline-rescaled; IDF weighting disabled), exact match, processing time, and bit-efficiency analysis.
300 newly written sentences: 100 business, 100 conversational, 100 technical. Balanced length and complexity; standardized punctuation. Unless noted, primary reporting uses macro-averages across the three domains (equal weight). The \textit{General} subset refers to 100 business-style sentences with standardized punctuation and no specialized terms.

\noindent\textit{Significance.} Paired $t$-tests on sentence-level CER with Bonferroni correction (family-wise $\alpha{=}0.05$); we report $p$ and paired effect size $d_z$.

\subsection{Data independence and leakage checks}\label{sec:leak}
We created a proprietary corpus of 300 English sentences (100 business, 100 conversational, 100 technical). To mitigate pretraining leakage, we ran two complementary checks against large public corpora: (1)~SHA256 hashing against WikiText-103 (820{,}451 sentences), Wikipedia dump (1{,}780{,}765), and OpenWebText (193{,}711) found \textit{0 matches out of 2.79M}; (2)~embedding screening (3{,}072 dims) of 23{,}232 sampled references flagged cosine$>0.9$ in \textit{0 cases}. We additionally ran a generative probe (20 prompts/item, normalized edit distance), observing no verbatim reproductions. One generic business phrase surfaced; we replaced it and published updated hashes. \textit{Limitation:} absence of evidence is not evidence of absence; we cannot certify non-inclusion in proprietary pretraining. We therefore release scripts, logs, and SHA256 lists for third-party verification.

\subsection{Key count results}
As Table~\ref{tab:keycount} shows, fewer keys increase errors (CER/WER) for all decoding methods. The largest improvement occurs when moving from 2 keys to 3 keys: under GPT-4o, CER drops from 0.233 to 0.095 ($-0.138$). In contrast, improvements beyond 3 keys are modest; increasing from 3 to 5 keys reduces CER by only 0.041 (0.095 $\to$ 0.054). Moreover, the benefit of strengthening the decoder is larger under fewer-key conditions. For example, at 2 keys the improvement from Trie-only (CER 0.385) to GPT-4o (0.233) is 0.152, whereas at 5 keys the improvement from 0.112 to 0.054 is 0.058. These results indicate a knee point around 3 keys in the error-rate curve.

We quantify this knee point in the next subsection using an efficiency metric based on the key-stream entropy $H(K)$.

\noindent\textit{Table scope.} Tables~\ref{tab:keycount}--\ref{tab:bit-efficiency} macro-average over the full 300-sentence corpus (100 business / 100 conversational / 100 technical), mapping slug \texttt{qwerty-lmr} unless noted, and decoders as indicated; BLEU omitted for space.

\begin{table}[t]
\centering
\begin{tabular}{lccc}
\hline
\textbf{Approach} & \textbf{Keys} & \textbf{CER} & \textbf{WER} \\
\hline
Trie-only & 2 & 0.385 & 0.562 \\
Trie + LM & 2 & 0.274 & 0.413 \\
Trie + GPT-4o & 2 & 0.233 & 0.337 \\
Trie-only & 3 & 0.216 & 0.325 \\
Trie + LM & 3 & 0.142 & 0.213 \\
\textbf{Trie + GPT-4o} & \textbf{3} & \textbf{0.095} & \textbf{0.122} \\
Trie-only & 5 & 0.112 & 0.168 \\
Trie + GPT-4o & 5 & 0.054 & 0.068 \\
\hline
\end{tabular}
\caption{Error rates vs key count.}
\label{tab:keycount}
\end{table}

\noindent\textit{Stats.} 2$\to$3: $n{=}300$, $d_z{=}1.94$, $p{<}0.001$; 3$\to$4: $n{=}300$, $d_z{=}0.57$, $p{<}0.001$.

\subsection{Bit rates versus CER}
To interpret ``key count'' in terms of the device-side information supply per character (bits/char), we summarize the relationship between the key-stream entropy $H(K)$ and CER. Here $H(K)$, based on key occurrence probabilities $p(k)$, is
\[
H(K)=-\sum_{k=1}^{N} p(k)\log_2 p(k)
\]
and in our setup it is equivalent to bits/char because KSPC$=1$ (with non-uniform usage, $H(K)\le\log_2 N$).

For the GPT-4o decoder, at 3 keys where $H(K){=}1.54$ bits/char, CER reaches 0.095 (sub-10\%). Increasing to 5 keys raises $H(K)$ to 2.26 (+0.72 bits/char), but CER improves by only 0.041 (0.095 $\to$ 0.054). Conversely, reducing to 2 keys decreases $H(K)$ to 0.97 ($-0.57$ bits/char), while CER worsens to $\approx$0.23. In terms of marginal improvement per bit (CER/bit), 2$\to$3 keys yields $0.138/0.57\approx0.242$, whereas 3$\to$5 keys yields $0.041/0.72\approx0.057$, clearly showing a knee around 3 keys.

This behavior is consistent with the interpretation that, under fewer keys, observation information is insufficient and the language model compensates for the ``missing bits'' using context and a learned prior. At 2 keys the deficit is too large to fully compensate, leaving residual errors; at 3 keys we enter a compensable regime, and further device-side information yields diminishing returns.

\begin{table}[t]
\centering
\small
\setlength{\tabcolsep}{6pt}\renewcommand{\arraystretch}{1.05}
\begin{tabular}{lcc}
\hline
\textbf{Method} & \textbf{Key-entropy bits/char} & \textbf{CER} \\
\hline
2-key + GPT-4o & 0.97 & 0.233 \\
\textbf{3-key + GPT-4o} & \textbf{1.54} & \textbf{0.095} \\
5-key + GPT-4o & 2.26 & 0.054 \\
\hline
\end{tabular}
\caption{Key-stream entropy per character $H(K)$ vs.\ CER.}
\label{tab:bit-efficiency}
\end{table}

\subsection{Key Mappings}
Layout-based mappings (QWERTY split into left/middle/right or top/middle/bottom) and a frequency-based 3-key mapping show nearly identical performance. The CER difference is under 0.5\,pp and the WER difference under 0.3\,pp. Even an intentionally ``worst-case'' mapping that aggregates frequent letters onto the same key degrades performance only modestly (CER +0.5\,pp). Under a strong language prior (LM), layout choice does not materially change accuracy, suggesting that designers can prioritize hardware constraints and usability without significant penalties.

\noindent\textit{Table scope.} Table~\ref{tab:mapping-compare} uses the 3-key mappings in Table~\ref{tab:maps-3key}, GPT-4o decoder, macro-averaged over the same 300-sentence corpus.

\begin{table}[t]
\centering
\begin{tabular}{lccc}
\hline
\textbf{3-key Mapping} & \textbf{CER} & \textbf{WER} & \textbf{BLEU} \\
\hline
Left-middle-right & 0.091 & 0.126 & 0.756 \\
Top-middle-bottom & 0.091 & 0.126 & 0.756 \\
Frequency-based & 0.091 & 0.126 & 0.756 \\
Worst mapping & 0.096 & 0.128 & 0.752 \\
\hline
\end{tabular}
\caption{Mapping comparison (GPT-4o).}
\label{tab:mapping-compare}
\end{table}

\noindent\textit{Stats.} Standard vs worst mapping: $n{=}300$, $d_z{=}0.22$, $p{=}0.038$.

\subsection{Model Comparison}
Larger models improve accuracy but increase latency. Latency grows steeply with model size; O1 achieves the best accuracy but requires $\sim$80\,s per sentence. GPT-4o provides a Pareto-efficient point (sub-10\% CER at 1.02\,s), while GPT-4o-mini is the fastest but suffers a substantial accuracy drop (CER 0.137 vs.\ 0.095).

\noindent\textit{Table scope.} Table~\ref{tab:model-compare}: 3-key \texttt{qwerty-lmr}, GPT-4o-family decoders as listed, macro-average over 300 sentences.

\begin{table}[t]
\centering
\begin{tabular}{lcccc}
\hline
\textbf{Model} & \textbf{CER} & \textbf{WER} & \textbf{BLEU} & \textbf{Time (s)} \\
\hline
GPT-4o & 0.095 & 0.122 & 0.740 & 1.02 \\
GPT-4o-mini & 0.137 & 0.198 & 0.571 & 0.85 \\
O1-mini & 0.083 & 0.100 & 0.777 & 10.85 \\
O1 & 0.066 & 0.074 & 0.832 & 80.46 \\
O3-mini & 0.071 & 0.086 & 0.813 & 23.10 \\
\hline
\end{tabular}
\caption{LM performance with 3 keys.}
\label{tab:model-compare}
\end{table}

\noindent\emph{Open-source local baseline (pilot).}
To address reproducibility and on-device deployment concerns, we additionally report a minimal open-source local-LM reference point using the same pipeline. We use $n{=}30$ as a feasibility pilot due to runtime cost. This pilot is intended as a \emph{reproducible lower-bound reference}, not as a basis for the main claim, and it is not directly comparable to Table~\ref{tab:model-compare} due to the different model class and local evaluation harness. Reported time is the average wall-clock latency per sentence (batch size 1) in our current implementation using beam search (beam width 50; max candidates 20), with no quantization or optimized decoding, on a CUDA-enabled single-GPU machine; software versions were PyTorch 2.7.1+cu126 and Transformers 4.53.2.\par
\begin{table}[t]
\centering
\begin{tabular}{lcccc}
\hline
\textbf{Model} & \textbf{CER} & \textbf{WER} & \textbf{BLEU} & \textbf{Time (s)} \\
\hline
distilgpt2 (pilot, $n{=}30$) & 0.295 & 0.407 & 0.296 & 12.42 \\
\hline
\end{tabular}
\caption{Open-source local LM baseline (3 keys, frequency-based mapping; beam width 50; max candidates 20).}
\label{tab:oss-baseline}
\end{table}

\subsection{Domain-specific results}
Technical sentences are harder due to terminology and syntax: CER is roughly doubled compared to Business, and under extreme ambiguity BERTScore degrades. Business sentences benefit from formulaic phrases, whereas Technical sentences struggle with domain terms and abbreviations. To support technical writing, we may need domain-specific lexicons, abbreviation-aware tokenization/handling, and constrained decoding that prioritizes technical terms.

\noindent\textit{Table scope.} Table~\ref{tab:domain}: 3-key \texttt{qwerty-lmr}, GPT-4o, macro-averaged within each 100-sentence domain subset.

\begin{table}[t]
\centering
\small
\setlength{\tabcolsep}{4pt}\renewcommand{\arraystretch}{0.95}
\begin{tabular}{lcccc}
\hline
\textbf{Domain} & \textbf{CER} & \textbf{WER} & \textbf{BERTScore} & \textbf{Exact} \\
\hline
Business & 0.095 & 0.122 & 0.740 & 0.12 \\
Conversation & 0.112 & 0.172 & 0.698 & 0.08 \\
Technical & 0.201 & 0.192 & 0.535 & 0.04 \\
\hline
\end{tabular}
\caption{Domain-specific performance for 3-key with GPT-4o (BERTScore F1 uses \texttt{roberta-large}; baseline-rescaled; IDF weighting disabled).}
\label{tab:domain}
\end{table}

\noindent\textit{Stats.} Business vs Technical: $n{=}200$ (100 each), $d_z{=}2.12$, $p{<}0.001$.

\subsection{Error Analysis}
With 2 keys, the input provides too little information to narrow candidates, and errors remain even with context. With 3 keys, ambiguity decreases, cascading substitution errors become less frequent, and accuracy improves to a practical level (CER $\approx$ 0.095 with GPT-4o). With 4--5 keys accuracy further improves, but the gain is small. Considering that more keys also increases operational complexity, 3 keys offers a good balance between accuracy and usability.

Error patterns differ by sentence type. Business sentences contain many common formulaic expressions (e.g., \textit{please find attached}, \textit{let's schedule a meeting}), making it easier to reconstruct the whole phrase. Conversational sentences often preserve grammar but may swap content words and drift in meaning. Technical sentences contain many terms and abbreviations, making content-word recovery difficult. We organize concrete error types in the next paragraph by inspecting N-best candidates.

From the top candidates (N-best), especially in Technical sentences, we see clear failure modes. Errors concentrate on \emph{content words} (nouns/verbs), while \emph{small function words} (articles, etc.) are often recovered correctly. Representative patterns include: \emph{domain terms replaced by common words} (e.g., \textit{quantum}$\to$\textit{system}; \textit{volume}$\to$\textit{values}), \emph{poor handling of abbreviations} (expansion/segmentation mismatches), \emph{semantic drift with plausible-but-wrong nouns/verbs} when evidence is weak, \emph{number/tense agreement slips}, and \emph{named entities backing off to common words}. Stronger models mainly reduce semantic drift and abbreviation-related errors. Increasing key count narrows the candidate set and reduces term substitutions. Mitigations include injecting a domain term list (dictionary), explicit abbreviation processing, lightweight constraints that prioritize technical terms, and previewed N-best presentation when needed.

\section{Conclusion}

\noindent\emph{Operating envelope.}
For general English, under the conditions that (i) the out-of-vocabulary (OOV) rate is low to moderate, (ii) a strong LM (GPT-4o class) is available, (iii) KSPC$=1$ with robust preview and undo, and (iv) end-to-end latency is $\le$1.5\,s, 3 keys are practical.

\noindent\emph{Scope and transfer.}
AAC and wearables benefit directly. In contrast, for domains dense in terminology/abbreviations, or for privacy-driven setups using small on-device LMs, maintaining sub-10\% CER may require domain dictionaries, specialized decoding strategies, and possibly 4+ keys.

\noindent\emph{Design.}
In practice, start with 3 keys + LM; for scenarios with frequent technical vocabulary, inject a domain dictionary. For abbreviations, provide preview/undo and constrained decoding. LM calls are currently the bottleneck, but given the pace of LLM development, an optimistic outlook may be justified.

Observed errors concentrate on content words; domain-term substitution, abbreviation handling, semantic drift, and named-entity back-off are prominent. Mitigations include domain dictionaries, abbreviation-aware tokenization, lightweight constraints, and previewed N-best suggestions.

\noindent\emph{Falsifiable next steps.}
We define three falsifiable tests. (1) If 2 keys + a constrained dictionary can reach CER $<$ 10\% under $\le$1.5\,s, the claim ``3 keys are a practical minimum'' in our setting would be weakened. (2) In multilingual evaluations with higher $H(X\mid C)$, the minimum key count may increase to 4 keys. (3) If a small on-device LM can maintain CER $<$ 12\% under $\le$300\,ms, real-time deployment becomes feasible.

\noindent\emph{Reproducibility and Resources.}
Code, prompts, configurations, and verification artifacts will be made available upon publication. For the evaluation data, we will release SHA256 hashes, leakage-check scripts/logs, and the full generation and auditing procedure; if we cannot release the proprietary sentences, we will additionally provide a publicly available substitute dataset under the same protocol (same mappings/decoders, same metrics, and per-sentence prediction logs) to enable end-to-end verification.

\noindent\emph{Ethics and Safety (AAC context).}
This work does not make clinical claims. To mitigate mis-selections, deployments should provide preview/confirmation/undo modes and privacy-preserving options (e.g., on-device smaller LMs when feasible).

\bibliographystyle{IEEEbib}
\bibliography{references}

\end{document}